\documentclass{ws-procs9x6}

\begin{document}

\title{ Phenomenology of Cosmic Ray Air Showers }

\author{Maria Teresa Dova\footnote{\uppercase{W}ork partially
supported by \uppercase{CONICET}, \uppercase{A}rgentina}}

\address{Departamento de F\'{\i}sica,\\ 
Universidad Nacional de La Plata, CC67 La Plata (1900), Argentina\\
E-mail: dova@fisica.unlp.edu.ar}

\maketitle

\abstracts{
The properties of cosmic rays with energies above $10^{6}$~GeV
have to be deduced from the spacetime structure and particle content of the air showers which
they initiate. In this review, a summary of the phenomenology of these
giant air showers is presented. We describe the  hadronic interaction
models used to extrapolate results from collider data to ultra high energies, an also the main electromagnetic processes that govern the longitudinal shower
evolution 
as well as the lateral spread of particles. 
}

\section{Introduction}

For primary cosmic ray energies above  $10^{6}$~GeV, the flux becomes so low that direct detection of the primary particle using detectors in or above the upper atmosphere is not longer possible. In these cases the primary particle has enough energy to initiate an extensive air shower (EAS) in the atmosphere. 
If the primary cosmic ray particle is a nucleon or nucleus the cascade begins with a hadronic interaction. The number of hadrons increases through subsequent generations of particle interactions. In each generation about 20\% of the energy is transferred to an electromagnetic cascade by rapid decays of neutral pions. Ultimately, the electromagnetic cascade dissipates roughly 90\% of the primary particle's energy trough ionization. The remaining energy is carried by muons and neutrinos from charged pion decays. The electromagnetic and weak interactions are rather well understood. However, uncertainties in hadronic interactions at ultra high energies
 constitute one of the most problematic sources of systematic error
 in analysis of air showers. In what follows a brief report of the phenomenology of EAS is presented, focusing in those aspects which are the main source of systematic uncertainties affecting somehow the determination of primary energy and mass composition. A complete review on the phenomenology of cosmic ray air showers can be found in ~\cite{Anchordoqui:2004xb}.

\section{Hadronic Processes}  
\label{hadronic}

Soft multiparticle production with small transverse momenta with respect to 
the collision axis is a dominant feature of most hadronic events at 
center-of-mass energies 
$10~{\rm GeV} < \sqrt{s} < 50~{\rm GeV}$. 
Despite the fact that strict calculations based on ordinary QCD perturbation 
theory are not feasible, there are some phenomenological models that 
successfully take into account the main properties of the soft diffractive 
processes. These models, inspired by $1/N$ QCD expansion are also 
supplemented with generally accepted theoretical principles like duality, 
unitarity, Regge behavior, and parton structure. The 
interactions are no longer described by single particle exchange, but by 
highly complicated modes known as Reggeons. Up to about 50~GeV, the slow growth of the cross 
section with $\sqrt{s}$ is driven by a dominant contribution of a special 
Reggeon, the Pomeron. 

At higher energies, semihard interactions arising from 
the hard scattering of partons that carry only a very small fraction of the 
momenta of their parent hadrons can compete successfully with soft 
processes. These semihard interactions lead 
to the minijet phenomenon, i.e., jets with transverse energy 
($E_T = |p_{_T}|$) 
much smaller than the total center-of-mass energy.  Unlike soft 
processes, this low-$p_{_T}$ jet physics can be computed in perturbative QCD. 
The parton-parton minijet cross section is given by
\begin{equation}
\sigma_{\rm QCD}(s,p_{{_T}}^{\rm cutoff}) = \sum_{i,j} \int 
\frac{dx_1}{x_1}\, \int \frac{dx_2}{x_2}\,
\int_{Q_{\rm min}^2}^{\hat{s}/2} \, d|\hat t|\,\, 
\frac{d\hat{\sigma}_{ij}}{d|\hat t|}\,\,
x_1 f_i(x_1, |\hat t|)\,\,\, x_2 f_j(x_2, |\hat t|) \,\,\,,
\label{sigmaminijet}
\end{equation}
where $x_1$ and $x_2$ are the fractions of the momenta of the parent hadrons 
carried by the partons which collide,
$d\hat{\sigma}_{ij}/d|\hat t|$ is the cross section for scattering of 
partons of types $i$ and $j$ according to elementary QCD diagrams, 
$f_i$ and $f_j$ are parton distribution functions (pdf's),
$\hat{s} = x_1\,x_2 s$ 
and $-\hat{t} = \hat{s}\, (1 - \cos \vartheta^*)/2 =  Q^2$ 
are the Mandelstam variables for this parton-parton process,
and the sum is over all parton species. The integration limits satisfy 
$Q_{\rm min}^2 < |\hat t| < \hat{s}/2,$ with $Q_{\rm min}$ the 
minimal momentum transfer. 

A first source of  uncertainty in modeling cosmic ray interactions at 
ultra high energy is encoded in the extrapolation of the measured 
parton densities  several orders of magnitude down to low $x$. 
Primary protons that impact on the upper atmosphere with energy $\approx~10^{11}$~GeV, 
yield partons with $x \equiv 2 p^*_{_\parallel}/\sqrt{s} \approx m_\pi/\sqrt{s} \sim  10^{-7},$ whereas current data on quark and gluon 
densities are only available for $x \approx 10^{-4}$ to within an experimental accuracy of 3\% for 
$Q^2 \approx 20$~GeV$^2$~\cite{Adloff:2000qk}. Moreover, application of HERA data to baryonic cosmic rays assumes universality
of the pdf's.

For large $Q^2$ and not too small $x$, the 
Dokshitzer-Gribov-Lipatov-Altarelli-Parisi (DGLAP) 
equations
successfully 
predict the $Q^2$ dependence of the quark and gluon densities ($q$ and $g,$ 
respectively).  In the double--leading--logarithmic 
approximation the 
DGLAP equations predict a steeply rising gluon density, $xg \sim x^{-0.4},$
which dominates the quark density at low $x$, 
in agreement with experimental results obtained with the HERA 
collider~\cite{Abramowicz:1998ii}.  Specifically, 
HERA data are found to be consistent with a 
power law, $xg(x,Q^2) \sim x^{-\Delta_{\rm H}},$ with an exponent $\Delta_{\rm H}$ 
between 0.3 and 0.4~\cite{Engel:ac}.

The high energy minijet cross section is then 
determined by the small-$x$ behavior of the parton distributions or, rather, 
by that of the dominant gluon distribution (via the lower limits of 
the $x_1$, $x_2$ 
integrations) which gives~\cite{Kwiecinski:1990tb}:
\begin{equation}
\sigma_{\rm QCD} (s)   
\propto \int_{2\,Q_{\rm min}^2/s}^{1} \frac{dx_1}{x_1}\,\,
x_1^{-\Delta_{\rm H}}\, \int_{2 \,Q_{\rm min}^2/s}^{1} \frac{dx_2}{x_2} 
\,\,x_2^{-\Delta_{\rm H}} \sim 
s^{\Delta_{\rm H}}\, \ln (s/s_0)\,,
\label{KH}
\end{equation}
where $s_0$ is a normalization constant. 
One caveat is that the inclusive QCD 
cross section given in Eq.~(\ref{KH}) is a Born approximation, and therefore 
automatically violates unitarity.

\begin{figure} [t]
\centerline{\epsfxsize=2.8in\epsfbox{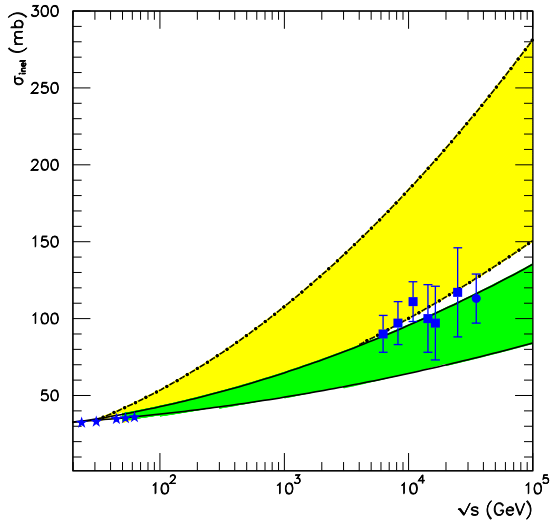}} 
\caption{Energy dependence of the $pp$ inelastic cross section as predicted by 
Eqs.~(\ref{qsig}) and (\ref{ssig})  with  $0.3 < \Delta_{\rm H} < 0.4.$
The 
darkly shaded region between the solid lines corresponds to the model 
with Gaussian parton  distribution in $\vec b$. The region between the 
dashed-dotted lines 
corresponds to the model with exponential fall-off of the parton density in $\vec b$.  
\label{sigmahera}}
\end{figure}

The procedure of calculating the inelastic cross section from inclusive cross 
sections is known as unitarization. In the eikonal 
model~\cite{Durand:cr} of high energy 
hadron-hadron scattering,the inelastic cross section, 
assuming a real eikonal function, is given by
\begin{equation}
\sigma_{\rm inel}=\int d^2\vec b\,
\left\{1-\exp\left[ -2\chi_{_{\rm soft}}(s,\vec b)
-2\chi_{_{\rm hard}}(s,\vec b)\right]\right\}\ ,
\label{inelastic}
\end{equation}
where the scattering is compounded as a sum of QCD ladders
via hard and soft processes through the
eikonals $\chi_{_{\rm hard}}$ and $\chi_{_{\rm soft}}$. It should be 
noted that we have ignored spin-dependent effects and the small real part of 
the scattering amplitude, both good approximations at high energies. Now, if the eikonal function, 
$\chi (s,\vec b) \equiv \chi_{_{\rm soft}}(s,\vec b) + 
\chi_{_{\rm hard}}(s,\vec b) =\lambda/2,$
indicates the mean number of partonic interaction pairs at impact parameter 
$\vec b,$ the probability $p_n$ for having $n$ independent partonic 
interactions using Poisson statistics reads, 
$p_n = (\lambda^n/n!) \, e^{-\lambda}$.
Therefore, the factor $1-e^{-2\chi} = \sum_{n=1}^\infty p_n$ in Eq.~(\ref{inelastic}) 
can be interpreted semiclassically as the probability 
that at least 1 of the 2 protons is broken up in a collision at impact 
parameter $\vec b$.
With this in mind, the inelastic cross section is simply the integral
over all collision impact parameters of the probability of having at 
least 1 interaction, yielding a mean minijet multiplicity of 
\mbox{$\langle n_{\rm jet} \rangle \approx \sigma_{\rm QCD}/
\sigma_{\rm inel}$~\cite{Gaisser:1988ra}.} The leading
contenders to approximate the (unknown) cross sections at
cosmic ray energies, {\sc sibyll}~\cite{Fletcher:1994bd} and 
{\sc qgsjet}~\cite{Kalmykov:te}, share the eikonal
approximation but differ in their {\em ans\"atse} for the
eikonals. In both cases, the core of dominant scattering at
very high energies is the parton-parton minijet cross section given in
Eq.~(\ref{sigmaminijet}),
\begin{equation}
\chi_{_{\rm hard}} = \frac{1}{2} \, 
\sigma_{\rm QCD}(s,p_{{_T}}^{\rm cutoff})\,\, A(s,\vec b) \,,
\label{hard}
\end{equation}
where the normalized profile function, $\int d^2\vec b \,\,A(s,\vec b) = 1,$ 
indicates the distribution of partons in the plane transverse to the 
collision axis. 

In the {\sc qgsjet}-like 
models, the  core of the hard eikonal 
is dressed with a soft-pomeron pre-evolution factor. This amounts to
taking a parton distribution which is Gaussian in the transverse
coordinate distance $|\vec b|.$ 

In  {\sc sibyll}-like models, the transverse density
distribution is taken as the Fourier transform of the proton electric
form factor, resulting in an energy-independent exponential 
(rather than Gaussian) fall-off of the parton density profile 
with $|\vec b|$. The main characteristics of the $pp$ cascade spectrum
resulting from these choices are readily predictable: the harder 
form of the {\sc sibyll}
form factor allows a greater retention of energy by the leading
particle, and hence less available for the ensuing 
shower. Consequently, on average {\sc sibyll}-like models predict a smaller  
multiplicity than {\sc qgsjet}-like models 
(see {\it e.g.}~\cite{Anchordoqui:1998nq,Anchordoqui:1999hn,Alvarez-Muniz:2002ne,Engel:is}). 

At high energy, $\chi_{_{\rm soft}} \ll \chi_{_{\rm hard}},$ and so the 
inelastic cross section is dominated by the hard eikonal. 
With the appropriate choice of 
normalization, the cross section in Eq.~(\ref{KH}) can be well--approximated 
by a power law. This implies 
that the growth of the inelastic  cross section according to {\sc qgsjet}-like models is given by
\begin{equation}
\sigma_{\rm inel} \sim \int d^2 \vec b \,\,\, \Theta (b_s - |\vec b|) = 
\pi b_s^2 \sim 4\pi \, \alpha^\prime_{\rm eff} \,\,\Delta_{\rm H}\,\,
\ln^2 (s/s_0) \sim 0.52 
\, \, \Delta_{\rm H} \,\, \ln^2 (s/s_0) \,\,\, {\rm mb}\,\,.
\label{qsig}
\end{equation}
For {\sc sibyll}-like models, the  growth of the inelastic
cross section also saturates the $\ln^2s$ Froissart 
bound,
but with a multiplicative constant which is larger 
than the one in {\sc qgsjet}-like models~\cite{Alvarez-Muniz:2002ne}. Namely,
\begin{equation}
\sigma_{\rm inel} \sim 3.2 \,\, \Delta_{\rm H}^2 \ln^2 (s/s_0) \,\,\, {\rm mb}\,\,.
\label{ssig}
\end{equation}

\begin{figure}[tbp]
\begin{minipage}[t]{0.49\textwidth}
\centerline{\epsfxsize=2.5in\epsfbox{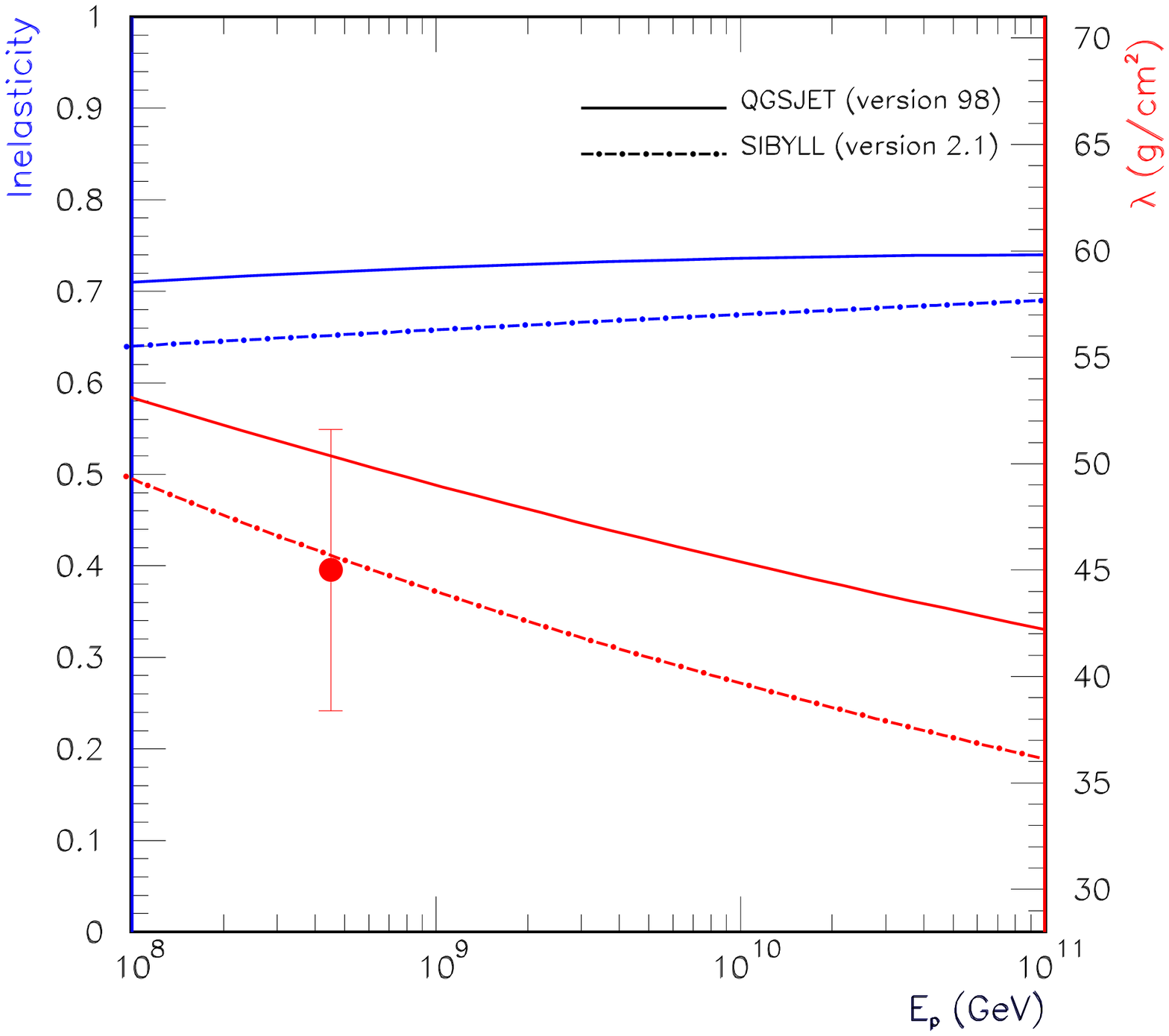}} 
\end{minipage}
\hfill
\begin{minipage}[t]{0.49\textwidth}
\centerline{\epsfxsize=2.22in\epsfbox{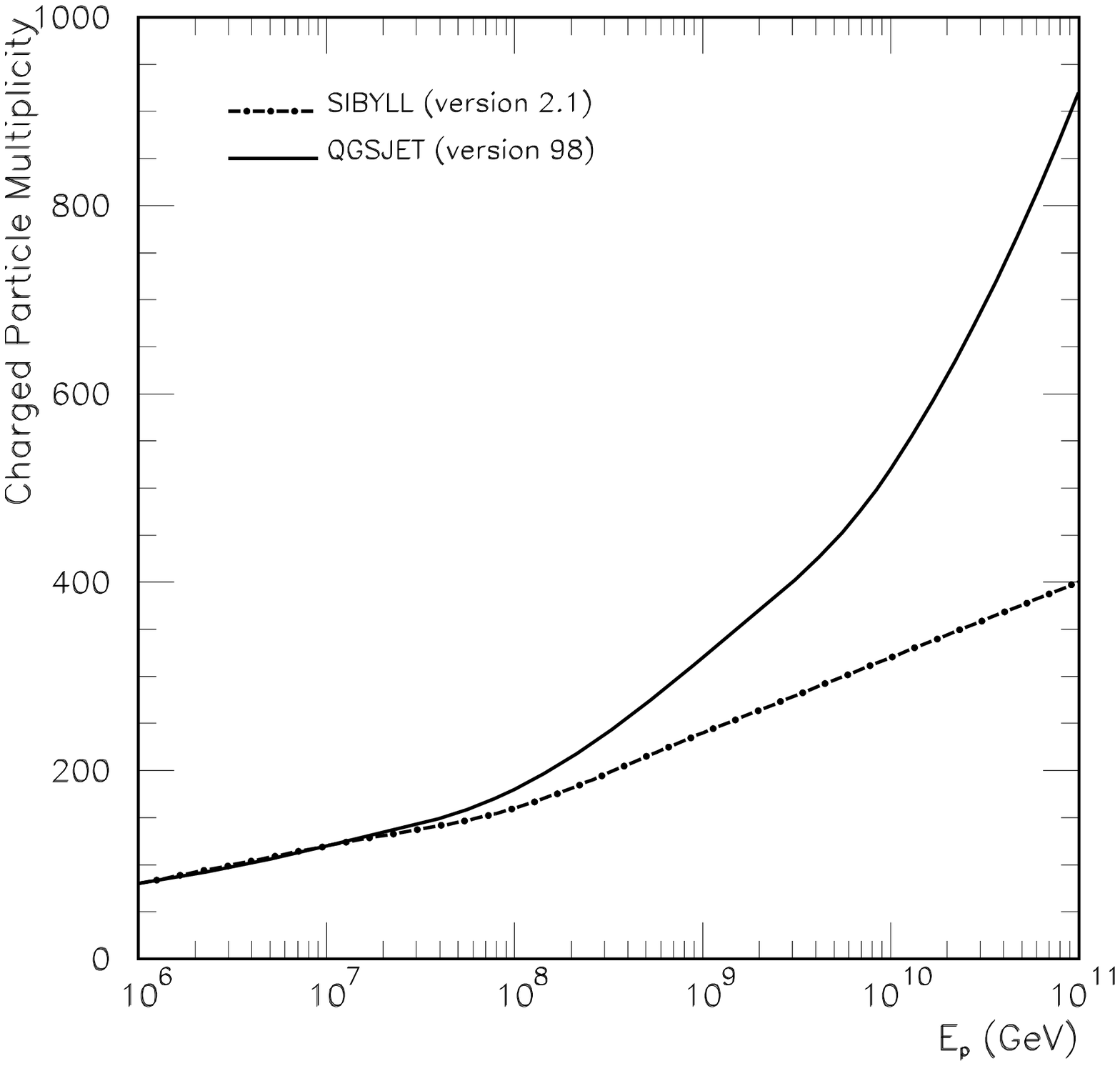}} 
\end{minipage}
\caption{Left panel:
The slowly rising curves indicate the mean 
inelasticity in  proton air collisions as predicted 
by {\sc qgsjet} and {\sc sibyll}. The falling 
curves indicate the proton mean free path in the atmosphere. 
Right panel: Mean multiplicity of charged secondary particles produced in inelastic 
proton-air collisions processed with {\sc qgsjet} and 
{\sc sibyll}.}
\label{mfp_n}
\end{figure}

Figure~\ref{sigmahera} illustrates the 
large range of predictions for $pp$ inelastic cross section which remain 
consistent with HERA data. When the two leading order approximations discussed above are
extrapolated to higher energies, both are consistent
with existing cosmic ray data.  Note, however, that in both cases the range of
allowed cross-sections at high energy varies by a factor of about 2 to 3.  The points in Fig.~\ref{sigmahera} 
correspond to the most up-to-date estimate of the $pp$ cross section from cosmic ray shower 
data ~\cite{Block:2000pg}.

There are three event generators, {\sc sibyll}~\cite{Fletcher:1994bd}, 
{\sc qgsjet}~\cite{Kalmykov:te}, 
and {\sc dpmjet}~\cite{Ranft:fd} 
which are tailored specifically for simulation of hadronic interactions up to 
the highest cosmic ray energies. The latest versions of these packages are {\sc sibyll} 
2.1~\cite{Engel:db}, {\sc qgsjet} 01~\cite{Heck01}, and {\sc dpmjet III}~\cite{Roesler:2000he}; 
respectively.  In {\sc qgsjet}, both the soft and hard 
processes are formulated in terms of Pomeron exchanges. To describe the 
minijets, the soft Pomeron  mutates into a ``semihard Pomeron'', 
an ordinary soft Pomeron with the middle piece replaced by a QCD parton 
ladder, as sketched in the previous paragraph.  This is generally referred to 
as the ``quasi-eikonal'' model.  In
contrast, {\sc sibyll} and {\sc dpmjet} follow a ``two channel'' eikonal model, where the soft and  
the semi-hard regimes are demarcated by a sharp cut in the transverse momentum: 
{\sc sibyll} 2.1 uses a cutoff parametrization inspired in the double leading logarithmic approximation 
of the DGLAP equations,
whereas {\sc dpmjet} III uses an {\it ad hoc} parametrization for the transverse momentum cutoff \cite{Engel:ac}.

The transition process from asymptotically free partons to colour-neutral 
hadrons is described in all codes by string fragmentation 
models~\cite{Sjostrand:1987xj}.  Different choices of fragmentation functions 
can lead to some differences in the hadron multiplicities. 
However, the main difference in the predictions of {\sc qgsjet}-like and 
{\sc sibyll}-like models arises from different assumptions in extrapolation of the parton 
distribution function to low energy.  

Now we turn to nucleus-nucleus interactions, which 
cause additional headaches for event generators 
which must somehow extrapolate $pp$ interactions
in order to simulate the proton-air collisions of interest. 
All the event generators described above adopt the
Glauber formalism~\cite{Glauber:1970jm}.  

Since the codes described above are still being refined, the disparity between
them can vary even from version to version.
At the end of the day, however, the relevant parameters 
boil down to two: the mean free path, 
$\lambda = (\widetilde \sigma_{\rm prod} \,n)^{-1},$ and the inelasticity,  
$K = 1 - E_{\rm lead}/E_{\rm proj}$, where 
 $n$ is the number density of atmospheric target nucleons,
$E_{\rm lead}$ is the 
energy of the most energetic hadron with a long lifetime, and $E_{\rm proj}$ 
is the energy of the projectile particle. Overall,
{\sc sibyll} has a shorter mean free path and a smaller inelasticity than 
{\sc qgsjet},
as indicated in Fig.~\ref{mfp_n}.  Since a shorter mean free path tends to 
compensate a smaller inelasticity, the two codes generate similar predictions 
for an air shower which has lived through several generations. The different 
predictions for the mean charged 
particle multiplicity in proton-air collisions are shown in Fig.~\ref{mfp_n}. 
Both models predict the same multiplicity below about $10^{7}$~GeV, but 
the predictions diverge above that energy. Such a divergence readily increases 
with rising energy. As it is 
extremely difficult to observe the first interactions experimentally,
it is not straightforward to determine which model is closer to reality.  

\section{Electromagnetic Component} 
\label{EM}

The evolution of an extensive air shower is dominated by electromagnetic
processes. The interaction of a 
baryonic cosmic ray  with an air nucleus high in the atmosphere leads to a 
cascade of secondary mesons and nucleons. The first few 
generations of charged pions interact again, producing a hadronic core, 
which continues to feed the electromagnetic and muonic components of the 
showers. Up to about $50$~km above sea level,
the density of atmospheric target nucleons is $n \sim 
10^{20}$~cm$^{-3},$ and so even for relatively
low energies, say  $E_{\pi^{\pm}}\approx 1$~TeV, the probability 
of decay before interaction falls below 10\%. 
Ultimately, the electromagnetic cascade dissipates around 90\%
of the primary particle's energy, and hence the total number of 
electromagnetic particles is very nearly proportional to the shower energy.

By the time a vertically incident 
$10^{11}$~GeV proton shower
reaches the ground, there are about $10^{11}$ secondaries with energy above
90~keV in the the annular region extending 8~m to 8~km from the shower core.
Of these, 99\%  are photons, electrons, and positrons, with a typical ratio 
of $\gamma$ to $e^+  e^-$ of 9 to 1. Their mean energy  is 
around 10~MeV and they transport 85\% of the total energy at ground level. Of course, 
photon-induced showers are even more dominated by the electromagnetic channel, 
as the only significant muon generation mechanism in this case is the decay of
charged pions and kaons produced in $\gamma$-air  interactions. 

It is worth mentioning that these figures dramatically change for the case of 
very inclined showers. For a primary zenith angle, $\theta > 70^{\circ},$ the 
electromagnetic component becomes attenuated exponentially with atmospheric 
depth, being almost completely absorbed at ground level. We remind the reader 
that the vertical atmosphere is $\approx 1000$ g/cm$^{2}$, and is about 36 
times deeper for completely horizontal showers.

In contrast to  hadronic collisions, the electromagnetic 
interactions of shower particles can be calculated very accurately from 
quantum electrodynamics. Electromagnetic interactions are thus not a 
major source of systematic errors in shower simulations. The first 
comprehensive treatment of electromagnetic showers was elaborated by 
Rossi and Greissen~\cite{Rossi}.  This treatment was recently cast in a more 
pedagogical form by Gaisser~\cite{Gaisser:vg} and a summary is presented in ~\cite{Anchordoqui:2004xb}.

The generation of the electromagnetic component is 
driven by electron bremsstrahlung and pair production~\cite{Bethe:1934za}.
Eventually the average energy per particle drops below a critical energy,
$\epsilon_0$, at which point ionization takes over from bremsstrahlung and pair
production as the dominant energy loss mechanism. The $e^\pm$ energy loss rate due to
bremsstrahlung radiation is nearly proportional to their energy, whereas the 
ionization loss rate varies only logarithmically with the $e^\pm$ energy.
 The changeover 
from radiation losses to ionization losses depopulates the shower.
One can thus categorize the shower development in three phases: the growth phase, in which all the particles 
have energy $> \epsilon_0$; the shower maximum, $X_{\rm max}$; and the shower 
tail, where the particles only lose energy, get absorbed or decay. 

The relevant quantities participating in the development
of the electromagnetic cascade are the probability for an electron of
energy $E$ to radiate a photon of energy $k=yE$ and
the probability for a photon to produce a pair $e^+e^-$
in which one of the particles (hereafter $e^-$) has energy $E=xk$.
These probabilities are determined by the properties of the air and 
the cross sections of the two processes. 

In the energy range of interest, the impact 
parameter of the electron or photon is larger than an atomic radius, so 
the nuclear field is screened by its electron cloud.  
In the case of complete screening, where the momentum transfer is small, 
the cross section for bremsstrahlung can be approximated by~\cite{Tsai:1973py}
\begin{equation}
\frac{d\sigma_{e \rightarrow \gamma}}{dk} \approx \frac{A_{\rm eff}}{X_0 N_A k}
\left(\frac{4}{3}-\frac{4}{3}y+y^2\right)\,\,,\label{brem}
\end{equation}
where $A_{\rm eff}$ is the effective mass number of the air, $X_0$ is a constant,
and $N_A$ is Avogadro's number. In the infrared limit ({\it i.e.}, $y \ll 1$) this approximation 
is inaccurate at the level of about 2.5\%, which is small compared to typical experimental 
errors associated with cosmic air shower detectors.  Of course, the approximation fails 
as $y \rightarrow 1$, when nuclear screening becomes incomplete, and as $y \rightarrow 0$, at which 
point the LPM and dielectric suppression effects become important.This 
infrared divergence is eliminated by the interference of 
bremsstrahlung amplitudes from multiple scattering centers.  
This collective effect of the electric potential of several
atoms is known as the Landau-Pomeranchuk-Migdal (LPM)
effect~\cite{Landau:um,Migdal:1956tc}.
Using similar approximations, the cross section
for pair production can be obtained~\cite{Tsai:1973py}.

The LPM suppression of the cross section results in an 
effective increase of the mean free path of electrons and photons. This 
effectively retards the development of the electromagnetic component of the shower. 
It is natural to introduce 
an energy scale, $E_{\rm LPM}$, at which the inelasticity is low enough that the LPM effect becomes 
significant~\cite{Stanev:au}.

The experimental confirmation of the LPM effect at Stanford Linear 
Accelerator Center (SLAC)~\cite{Klein:1998du} has motivated new analyses of 
its consequences in cosmic ray 
physics~\cite{Alvarez-Muniz:1998px,Cillis:1998hf,Capdevielle:jt,Plyasheshnikov:2001xw}.
The most evident signatures of the LPM effect on shower development are  
a shift in the position of the shower maximum $X_{\rm max}$ and larger 
fluctuations in the shower development.

\begin{figure}[tbp]
\begin{minipage}[t]{0.49\textwidth}
\centerline{\epsfxsize=2.4in\epsfbox{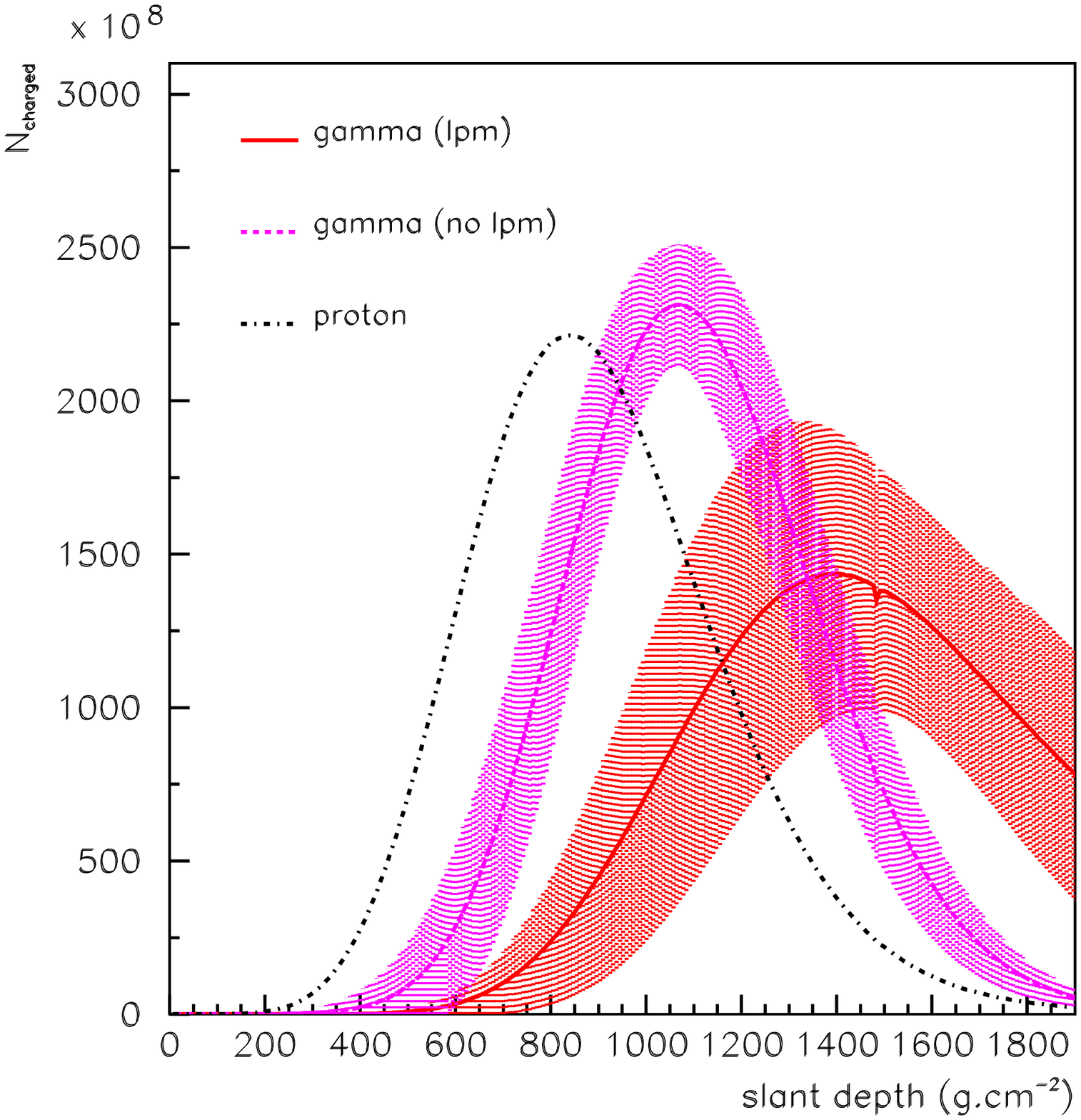}} 
\end{minipage}
\hfill
\begin{minipage}[t]{0.49\textwidth}
\centerline{\epsfxsize=2.5in\epsfbox{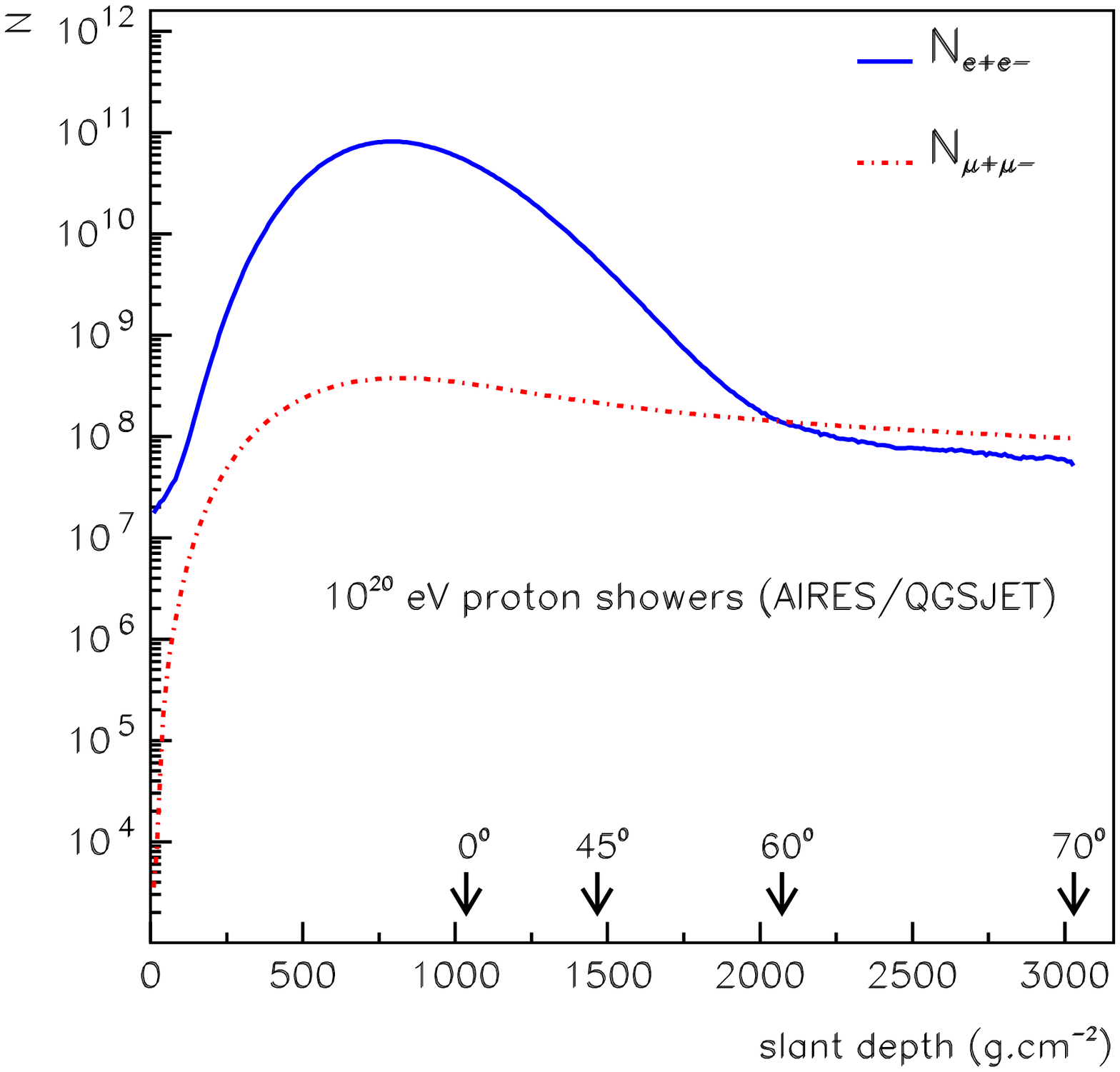}} 
\end{minipage}
\caption{Left panel:Average longitudinal shower developments 
of  $10^{11}$~GeV proton (dashed-dotted line) and $\gamma$-rays with 
and without the LPM effect (solid and dotted lines, respectively). The 
primary zenith angle was set to $\theta = 60^{\circ}$.
Right panel: Longitudinal development of  muons and electrons as a function of the slant depth 
for $10^{11}$~GeV proton-induced showers.}
\label{longi}
\end{figure}

Since the upper atmosphere is very thin the 
LPM effect becomes noticeable only for photons and electrons with energies above
$E_{\rm LPM} \sim 10^{10}$~GeV.  For baryonic primaries the LPM effect does not 
become important until the primary energy exceeds $10^{12}$GeV.  
To give a visual impression of how the LPM effect 
slows down the initial growth of high energy photon-induced showers,
we show the average longitudinal 
shower development of  $10^{10}$~GeV proton and $\gamma$-ray showers (generated using 
{\sc aires} 2.6.0~\cite{Sciutto:1999jh}) with and without the LPM effect in Fig.~\ref{longi}.

At energies at which the LPM effect is important ({\it viz.}, $E > E_{\rm LPM}$), 
$\gamma$-ray showers will have already commenced in the geomagnetic field at almost 
all latitudes.  This reduces 
the energies of the primaries that reach the atmosphere, and thereby  
compensates the tendency of the LPM effect to retard the shower development.
The first description of photon interactions in the geomagnetic field 
dates back at least as far as 1966~\cite{Erber:1966vv},
with a punctuated revival of activity in the early 1980's~\cite{Mcbreen:yc}.
More recently, a rekindling of interest in the topic has led to 
refined calculations~\cite{Bertou,Homola:2003ru}.   
Primary photons with energies above  $10^{10}$~GeV convert into $e^{+}e^{-}$ 
pairs, which in turn emit synchrotron photons. Regardless of the primary energy, 
the spectrum of the resulting 
photon ``preshower'' entering the upper atmosphere 
extends over several decades below the primary photon energy, and is peaked 
at energies below  $10^{10}$~GeV~\cite{Bertou}.  The geomagnetic cooling
thus switches on at about the same energy at which the LPM effect does, and 
thereby preempts the LPM-related observables which would otherwise be evident.
 
The relevant parameter to 
determine both conversion probability and synchrotron emission  is 
$E\times B_{\perp}$, where $E$ is the $\gamma$-ray energy and $B_{\perp}$ the transverse 
magnetic field. This leads to a large 
directional and geographical dependence of shower observables. Thus, each experiment has its own 
preferred direction for identifying primary gamma rays. 

\subsubsection{Electron lateral distribution function}

The transverse development of electromagnetic showers is dominated by Coulomb 
scattering of charged particles off the nuclei in the atmosphere. The lateral 
development in electromagnetic cascades in different materials 
scales well with the Moli\`ere radius $r_{\rm M} = E_s\, X_0/\epsilon_0,$ which varies
inversely with the density of the medium,$ 
r_{\rm M}  = r_{\rm M}(h_{\rm OL}) \,\,\frac{\rho_{\rm atm}(h_{\rm OL})}{\rho_{\rm atm}(h)} \simeq 
\frac{9.0~{\rm g}/{\rm cm}^{2}}{\rho_{\rm atm}(h) } $, 
where $E_s \approx 21$~MeV and the subscript ${\rm OL}$ indicates 
a quantity taken at a given observation level.

Approximate calculations of cascade equations in three dimensions to derive 
the lateral structure function for a pure electromagnetic cascade in vertical showers were 
obtained by Nishimura and Kamata~\cite{Kamata}, and later worked out by Greisen~\cite{Greisen} 
in the well-known NKG formula,
\begin{equation}
\rho_{}(r)  =  \frac{N_e}{r_{\rm M}^2} \, C \, \left( \frac{r}{r_{\rm M}}
\right)^{s_{_{\rm NKG}} -2} \left(1+\frac{r}{r_{\rm M}}\right) ^{s_{_{\rm NKG}}-4.5} \,\,,
 \label{nkg}
\end{equation}
where $N_e$ is the total number of electrons, $r$ is the distance from the shower axis.
For a primary of energy $E_0,$ the  so-called ``age parameter'', $s_{_{\rm NKG}} = 3\, /(1+\frac{2\, \ln(E_0/\epsilon_0)}{t} )$,
characterizes the stage of the shower development in terms of 
the depth of the shower in radiation lengths, {\it i.e.},
$t = \int_{z}^{\infty} \rho_{\rm atm}(z) \,\,dz/X_0.$

The NKG formula may also be extended to describe showers initiated
by baryons~\cite{Dova:2001jy}.  In such an extension, one finds 
a deviation of behavior of the Moli\`ere radius when using a 
value of the age parameter which is derived from theoretical 
predictions for pure electromagnetic cascades.
It is possible to generalize the NKG formula for the electromagnetic component of 
baryon-induced showers by modifying the exponents in Eq.~(\ref{nkg})~\cite{Dova:2001jy}. The derived NKG formula provides a good description of 
the $e^{+}e^{-}$ lateral distribution 
at all stages of shower development for values of $r$
sufficiently far from the hadronic core.
Fortunately, this is the experimentally interesting region, since typical
ground arrays can only measure densities at 
$r > 100 $~m from the shower axis, where detectors are not saturated.
 It should be mentioned that an NKG-like formula can be used to 
parametrize the total particle's density observed in baryon-induced showers~\cite{Roth:2003rs}.

In the case of inclined showers, one normally analyzes particle densities
in the plane perpendicular to the shower axis.  Simply projecting distributions
measured at the ground into this plane is a reasonable approach for near-vertical showers, but
is not sufficient for inclined showers.   In the latter case, 
additionally asymmetry is introduced because of both 
unequal attenuation of the electromagnetic components arriving at the ground earlier than and 
later than the core~\cite{Dova:2001jy}. Moreover, deflections on the geomagnetic field become important for showers 
inclined by more than about $70^\circ.$

In the framework of cascade theory, any effect coming from the influence of the
atmosphere should be accounted as a function of the slant depth 
$t$~\cite{Kamata}. Following this idea, a LDF valid 
at all zenith angles $\theta < 70^{\circ}$ can be determined by considering
\begin{equation}
t'(\theta,\zeta)  =  t\,\sec\theta\,(1+ K\,\cos\zeta)^{-1}\,,
\end{equation}
where $\zeta$ is the azimuthal angle in the shower plane, 
$ K =   K_0\,\tan\theta,$ and $K_0$ is a constant extracted 
from the fit~\cite{Dova:2001jy,Dova:dq}.  Then, the  particle lateral distributions 
for inclined showers $\rho(r,t')$ are  given by 
the corresponding vertical LDF $\rho (r,t)$ but evaluated at slant 
depth $t'(\theta,\zeta)$ where the dependence on the azimuthal angle is evident.

For zenith angles $\theta > 70^{\circ}$, the surviving electromagnetic component at ground is 
mainly due to muon decay and, to a much smaller extent, hadronic interactions, pair production 
and bremsstrahlung. As a result the lateral distribution follows that of the muon rather closely.
In Fig.~\ref{longi} the longitudinal development of the muon and electron components are shown. It is 
evident from the figure that  for very inclined showers  
the electromagnetic development is due mostly to muon decay~\cite{Ave:2000dd,Cillis:ij}. 

\subsection{The muon component}

The muonic component of EAS differs from the electromagnetic component for two main reasons.
First, muons are generated through the decay of cooled  charged pions, and thus the muon content is sensitive to the initial
baryonic content of the primary particle.  Furthermore, since there is no ``muonic cascade'', the number
of muons reaching the ground is much smaller than the number of electrons. Specifically, there are
about $5\times 10^{8}$ muons above 10~MeV at ground level for a vertical $10^{11}$~GeV proton 
induced shower.
Second,  the muon has a much smaller cross section for radiation and pair production than the electron, 
and so the muonic component of EAS develops differently than does the electromagnetic component.
The smaller multiple scattering suffered by muons leads to 
earlier arrival times at the ground for muons than for the electromagnetic component.  

The ratio
of electrons to muons depends strongly on the distance from the core;
for example, the $e^+  e^-$ to  $\mu^+  \mu^-$ ratio for a $10^{11}$~GeV vertical proton 
shower varies from 17 to 1 at 200~m from the core to 1 to 1 at 2000~m.
The ratio between the electromagnetic and muonic shower components behaves somewhat differently
in the case of inclined showers.  For zenith angles greater than $60^{\circ}$, the 
$e^+  e^-$/$\mu^+  \mu^-$  ratio  
remains roughly constant at a given distance from the core.  As the zenith angle grows beyond $60^\circ$,
this ratio decreases, until at $\theta = 75^{\circ}$, it is 400 times smaller than for a vertical shower.
Another difference between inclined and vertical showers is that the average muon energy at ground changes 
dramatically.  For horizontal showers, the lower energy muons are filtered out by a combination of 
energy loss mechanisms and the finite muon lifetime: for vertical showers, the average muon energy 
is 1~GeV, while for horizontal showers it is about 2 orders of magnitude greater.

High energy muons lose energy through $e^{+}e^{-}$ pair production, 
muon-nucleus 
interaction, bremsstrahlung, and knock-on electron ($\delta$-ray) 
production~\cite{Cillis:2000xc}.  The first three processes 
are discrete in the sense that they are characterized by high inelasticity and a large mean free path.  
On the other hand,  because of its 
short mean free path and its small inelasticity, knock-on electron production can be 
considered a continuous process. The muon bremsstrahlung cross section is suppressed by a factor of 
$(m_e / m_\mu)^2$ with respect to electron bremsstrahlung, see Eq.~(\ref{brem}).  Since the radiation 
length for 
air is about $36.7~\mathrm{g}/\mathrm{cm}^2$, and the vertical atmospheric depth is 1000~g/cm$^2$, 
muon bremsstrahlung is of negligible importance for vertical air shower development.  Energy loss due to 
muon-nucleus interactions is somewhat smaller than muon bremsstrahlung.    
Energy  loss by pair production is slightly more important than 
bremsstrahlung at about 1~GeV, and becomes increasingly dominant with energy.  
Finally, knock-on electrons have a 
very small mean free path, but also a very small inelasticity, so that this 
contribution to the energy loss is comparable to that from the hard processes.

In addition to muon production through charged pion decay, photons can directly generate muon pairs,
or produce hadron pairs which in turn decay to muons. In the case of direct pair production, the large 
muon mass leads to a higher threshold for this process than for electron pair production. 
Furthermore, QED predicts that $\mu^+ \mu^-$ production is suppressed by a factor 
$(m_{e}/m_{\mu})^2$ compared the Bethe-Heitler cross section. The cross section for hadron production by 
photons is much less certain, since it involves the {\it hadronic structure} of the photon. This has been 
measured at HERA for photon energies corresponding to $E_{ \mathrm{lab}} = 2 
\times 10^{4}$~GeV.  This energy is still well below the energies of 
the highest energy cosmic rays, but nonetheless, these data do constrain the extrapolation of the cross 
sections to high energies. 

The muon content of the shower tail is quite sensitive to unknown details of 
hadronic physics.  This implies that attempts to extract composition information from measurements
of muon content at ground level tend to be systematics dominated. 
The muon LDF is mostly determined by the distribution in phase space of the parent
pions.  However, the pionization process together with muon deflection in the geomagnetic field obscures 
the distribution of the first generation of pions. 
A combination of detailed simulations, high
statistics measurements of the muon LDF  and
identification of the primary species using uncorrelated observables 
could shed light on hadronic interaction models. 

\section*{Acknowledgments}

I would like to thank the organizers for the financial support and warn hospitality. I am grateful to L. Anchordoqui, A. Mariazzi, T. McCauley, T. Paul, S. Reucroft and J Swain for providing a very productive and agreeable working atmosphere in order to write our review article.

\end{document}